\documentclass[slac_one]{revtex4}
\usepackage{graphicx}
\usepackage{fancyhdr}
\newcommand{\rparity}{$R_p$}
\newcommand{\MET}{\mbox{${\hbox{$E$\kern-0.6em\lower-.1ex\hbox{/}}}_T$}}

\begin{document}

\title{{\small{Hadron Collider Physics Symposium (HCP2008),
Galena, Illinois, USA}}\\ 
\vspace{12pt}
SUSY Searches at the Tevatron} 

%

\author{T. Adams, {\it for the CDF and D\O\ Collaborations}}
\affiliation{Florida State University, Tallahassee, FL 32306, USA}

\begin{abstract}
Numerous searches for evidence of supersymmetry at the Tevatron
have been performed by the CDF and D\O\ collaborations.  Recent
results are presented here including squarks and gluinos
in jets and missing transverse energy, stop in several
different decay modes, charginos and neutralinos in trileptons,
neutralinos in di-photons, R-parity violating sneutrinos in
e+$\mu$ events, and long-lived particles.  These explore
many variations of SUSY such as MSSM, mSUGRA, and GMSB.
While no evidence
of SUSY production is observed, 95\% CL limits on cross sections 
and SUSY parameter space are set.  Most of these
limits are the world's best.
\end{abstract}

\maketitle

\thispagestyle{fancy}


\section{INTRODUCTION}

Of the many Beyond the Standard Model (BSM) theories,
supersymmetry (SUSY) is one of the most investigated
models.  SUSY provides a possible solution to the
hierarchy problem~\cite{bib:susyprimer} and is a
necessary component of string theory.
SUSY proposes a new symmetry between fermions
and bosons that doubles the number of particles.
This symmetry introduces a new multiplicative quantum number,
$R$-parity (\rparity).  Supersymmetric particles have
$R_p=-1$, while normal (non-SUSY) particles
have $R_p=+1$.

SUSY comes in many varieties (see M.~Perelstein's
contribution to these proceedings for a review of
recent theoretical work).  The mechanism responsible
for supersymmetry breaking creates many of these
variations including mSUGRA (super-gravity),
GMSB (gauge-mediated supersymmetry breaking), and
AMSB (anomaly-mediated supersymmetry breaking).
If \rparity\ is conserved then SUSY particles must
be produced in pairs at the Tevatron and cascade down
to the lightest supersymmetric particle (LSP) which
is stable.  The LSP then becomes a dark matter
candidate, simultaneously solving problems in particle
physics and astronomy.  

At the Fermilab Tevatron, supersymmetry has been searched
for in a wide variety of channels, some that complement
previous LEP searches and some that explore completely
new territory.  Here we present many of the latest
results.
This proceedings is divided into two categories which
are label ``natural'' SUSY (Sec.~\ref{sec:natural})
and ``unnatural'' SUSY (Sec.~\ref{sec:unnatural}).  
These are labels applied by the author and not intended
to be used as formal definitions.  All limits quoted
are at the 95\% CL.  For a complete review of Tevatron
Run II searches (including Higgs) through the first seven 
years of running see~\cite{bib:arnaudreview}.

\section{``NATURAL'' SUSY \label{sec:natural}} 

What is categorized here as ``natural'' SUSY are models
where \rparity\ is conserved and all SUSY particles (other
than the LSP) promptly decay.  Since astronomical bounds
require a stable LSP to be neutral~\cite{bib:astrostable},
LSP candidates are generally the lightest neutralino 
($\tilde{\chi_1^0}$), the sneutrino ($\tilde{\nu}$) or
the gravitino ($\tilde{G}$).  The prompt decays ensure that
all particles observed directly in the detector are 
non-supersymmetric and originate from the production
vertex.  The LSP will escape
the detector without interacting resulting in a missing
transverse energy signature.

\subsection{Squarks and Gluinos} 

\begin{figure}
 \includegraphics[width=0.9\textwidth]{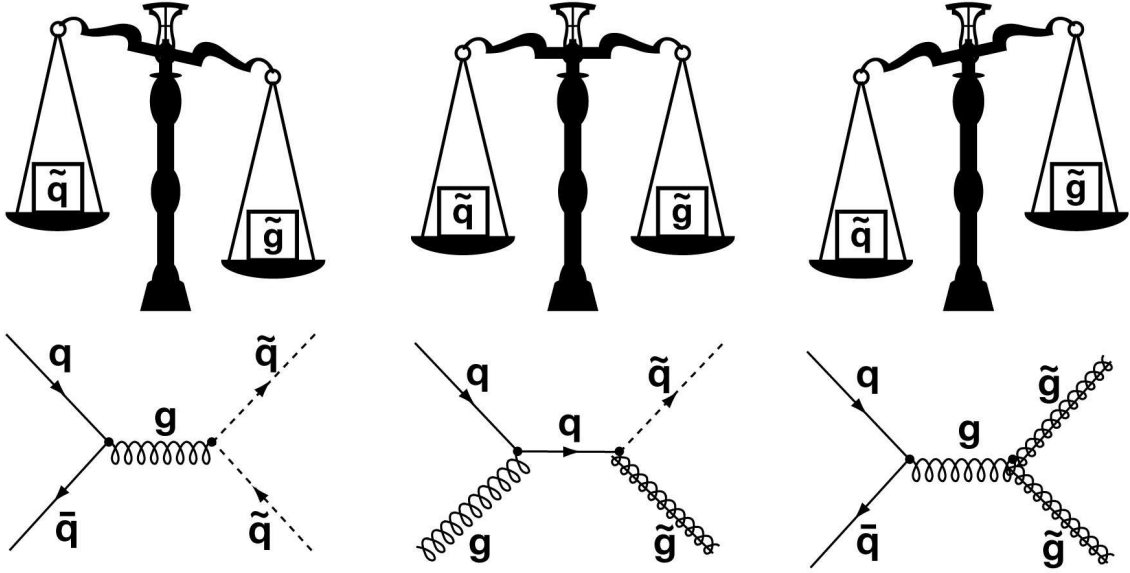}%
 \caption{Squark and gluino production processes for various
   mass relations:  M(squark) $\ll$ M(gluino) (left), 
   M(squark) $\approx$ M(gluino) (middle), M(squark) $\gg$ M(gluino)
   (right).  \label{fig:scales}}
\end{figure}

At hadron colliders (such as the Tevatron), strong interactions
tend to dominate production processes.  Therefore, 
squarks and gluinos may have the highest production rates,
if their masses are not too much larger than other SUSY particles.
Then if the squark mass is significantly less than the gluino, 
squark pair production dominates.  In the opposite case, 
gluino pair production is largest.  Finally, if the masses
are similar, squark+gluino production becomes important 
(see Fig.~\ref{fig:scales}).

The primary decay modes are $\tilde{q} \rightarrow q\tilde{\chi}_1^0$
and $\tilde{g} \rightarrow q\bar{q}\tilde{\chi}_0^1$.  Therefore,
each of these mass conditions leads to a different experimental
signature of two to four primary jets from decays.  Additional
jets can be found in the event from mis-reconstruction or
initial/final state radiation.  The
generic signature is multiple jets with missing transverse
energy (\MET).  In order to optimize sensitivity three
concurrent analyses are carried out looking for (a) two or
more jets, (b) three or more jets, and (c) four or more jets.
Events that fall into more than one sample are accounted for
in the combination.

\begin{figure}
 \includegraphics[width=0.45\textwidth]{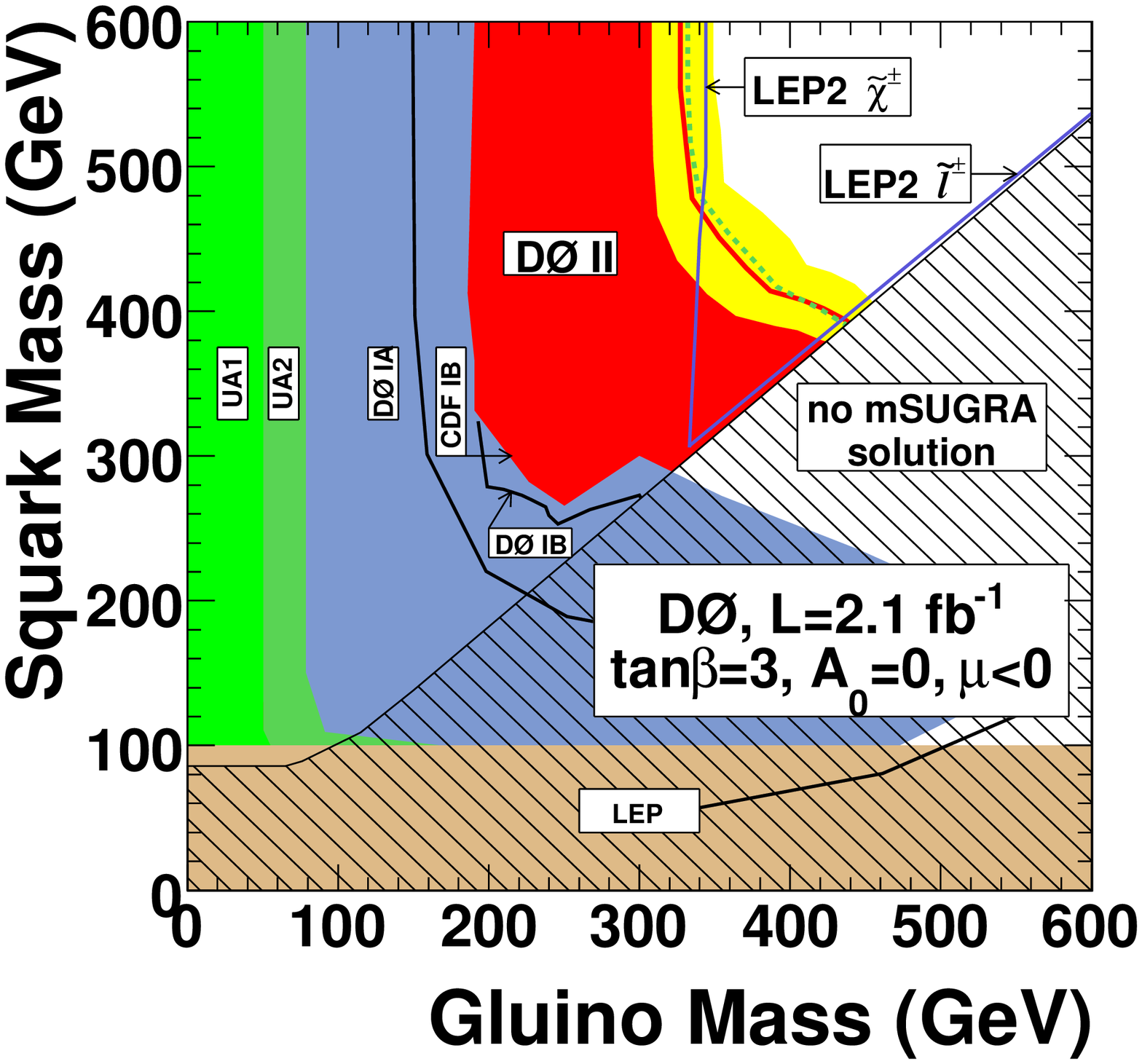}%
 \includegraphics[width=0.45\textwidth]{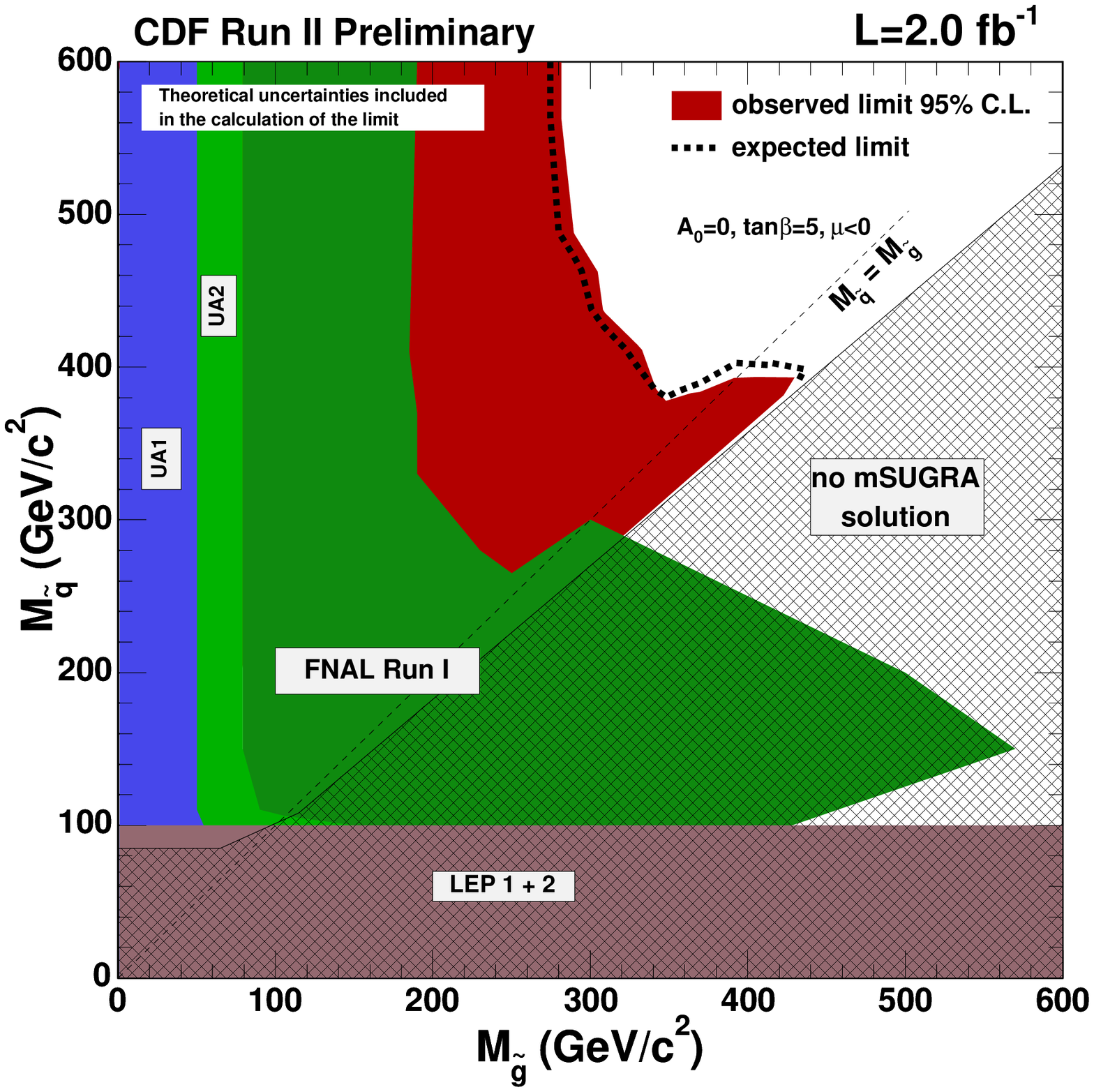}%
 \caption{Limits on the squark and gluino masses from D\O\ (left)
   and CDF (right). \label{fig:sqgl}}
\end{figure}

Both CDF and D\O\ have performed searches for squarks and
gluinos with more than 2 fb$^{-1}$ of Run II 
data~\cite{bib:cdfsqgl,bib:d0sqgl}.  Events
are selected with multiple high-$p_T$ jets and large
missing transverse energy.  The mSUGRA model is used
with $A_0 = 0$, $\mu < 0$, and $tan\beta = 3$ for D\O\ and
5 for CDF to simulate
signal.  For each of the jet multiplicities, the analyses
are optimized for final selection on jet $p_T$, \MET,
and total energy.  No significant excess of data is observed.
Figure~\ref{fig:sqgl} shows the interpretation of the
analysis as limits on the squark versus gluino mass 
plane within the mSUGRA model.

The large top quark mass contributes to the stop mass to generally 
cause it to be the lowest mass squark.  It also causes
the $\tilde{t} \rightarrow t\tilde{\chi}_1^0$ decay to be 
suppressed, therefore allowing other decay modes of 
interest.  Therefore each collaboration has performed
dedicated stop searches.

\begin{figure}
 \includegraphics[width=0.55\textwidth]{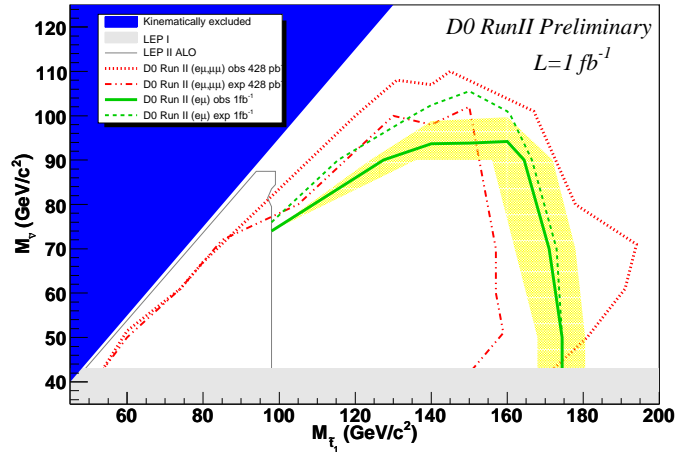}%
 \caption{Limits on the stop and sneutrino masses from the D\O\ 
   stop search in the $e + \mu + $ \MET\ channel.\label{fig:d0stopemu}}
\end{figure}

The D\O\ collaboration has searched for stop in events
with an electron, a muon, two or more jets and large missing
transverse energy~\cite{bib:d0stopemu}.  The different
type leptons leads to a higher branching ratio and lower 
backgrounds in a search for the decay 
$\tilde{t} \rightarrow \ell b \tilde{\nu}$ where
$\tilde{\nu}$ is the LSP.  The selected events were binned
two dimensionally in the variables $\Sigma p_T (jets)$
and $p_T (\mu) + p_T (e) + $\MET.  Comparisons of data and
expected backgrounds show good agreement.  Limits are
set in the $m_{\tilde{\nu}}$ vs $m_{\tilde{t}}$ parameter
space (Fig.~\ref{fig:d0stopemu}).

CDF has performed a similar search in the dilepton channel
(lepton = $e$ or $\mu$) with the decay mode
$\tilde{t} \rightarrow b \tilde{\chi}_1^\pm \rightarrow 
b \ell \tilde{\chi}_1^0 \nu$ where the $\tilde{\chi}_1^0$ is the 
LSP~\cite{bib:cdfstopbchi}.  This search is targeted for stop
masses below the top mass ($m_{\tilde{t}} = 135-155$ GeV).
Good agreement is observed between data and expected
background.  However, under the assumption that the decay 
chain is dominated by 
$\tilde{\chi}_1^\pm \rightarrow W^\pm \tilde{\chi}_1^0$
no limits on stop production are possible.  (Note, this
analysis has been updated to higher luminosity and to 
allow for higher branching fractions $BR(\tilde{\chi}_1^\pm
\rightarrow b \ell \tilde{\chi}_1^0 \nu)$~\cite{bib:cdfstopbchi}.

\begin{figure}
 \includegraphics[width=0.55\textwidth]{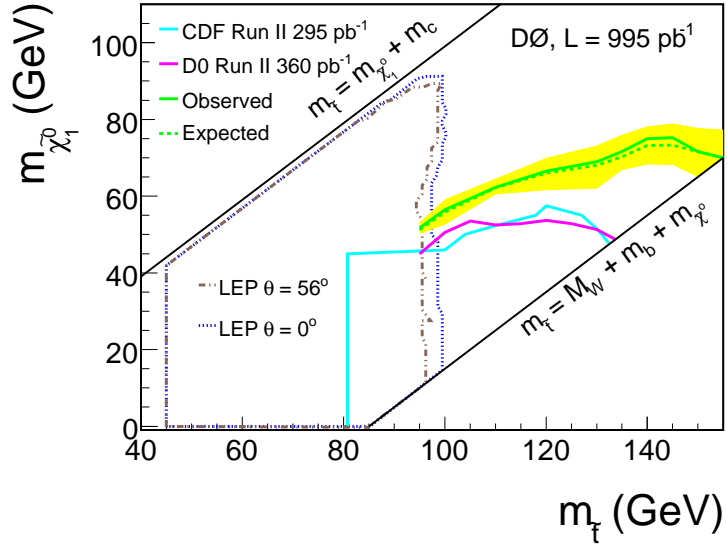}%
 \caption{Limits on stop and neutralino masses from the D\O\ stop search
   in the charm + MET channel.  \label{fig:d0stopcharm}}
\end{figure}

Another stop search by D\O\ looked for the decay mode $\tilde{t} \rightarrow
c\tilde{\chi}_1^0$~\cite{bib:d0ccmet}.  If the mass relations 
$m_{\tilde{t}} < m_t + m_{\tilde{\chi}_1^0}$,
$m_{\tilde{t}} < m_{b} + m_{\tilde{\chi}_1^\pm}$ and
$m_{\tilde{t}} < m_W + m_b + m_{\tilde{\chi}_1^0}$ are true,
then the flavor-changing charm + \MET\ is assumed to
occur with 100\% branching ratio.  The analysis selected
events with two jets (one of which is tagged as heavy
flavor) and large \MET.  Final selections on \MET,
$S = \Delta\phi_{max} + \Delta\phi_{min}$, and the
scalar sum of the $p_T$ of all jets ($H_T$) are 
optimized for three different stop and neutralino masses.
Good agreement between data and background leads to
limits in the $m_{\tilde{\chi}_1^0}$ vs $m_{\tilde{t}}$
plane (Fig.~\ref{fig:d0stopcharm}).

\begin{figure}
 \includegraphics[width=0.45\textwidth]{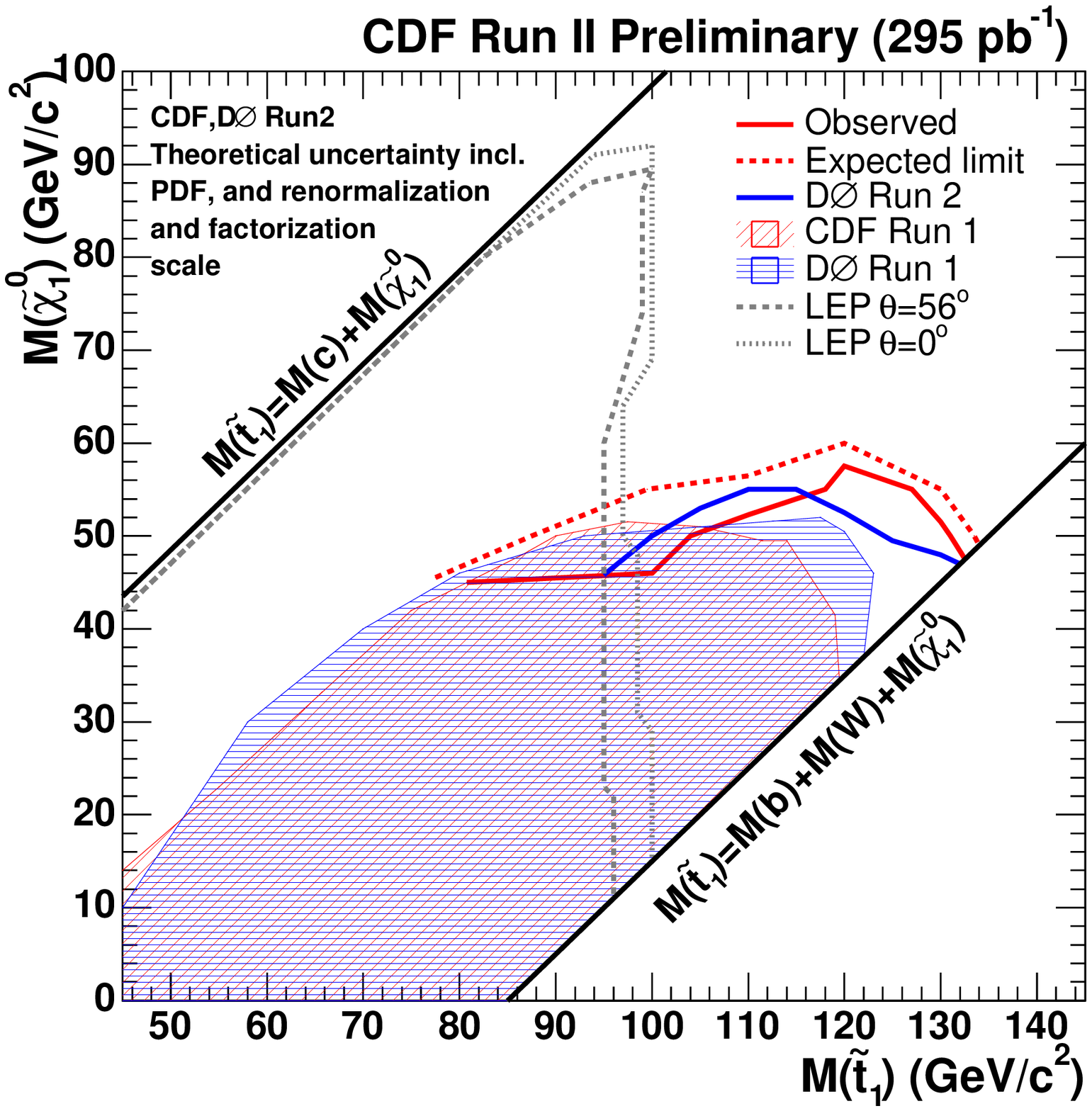}%
 \includegraphics[width=0.45\textwidth]{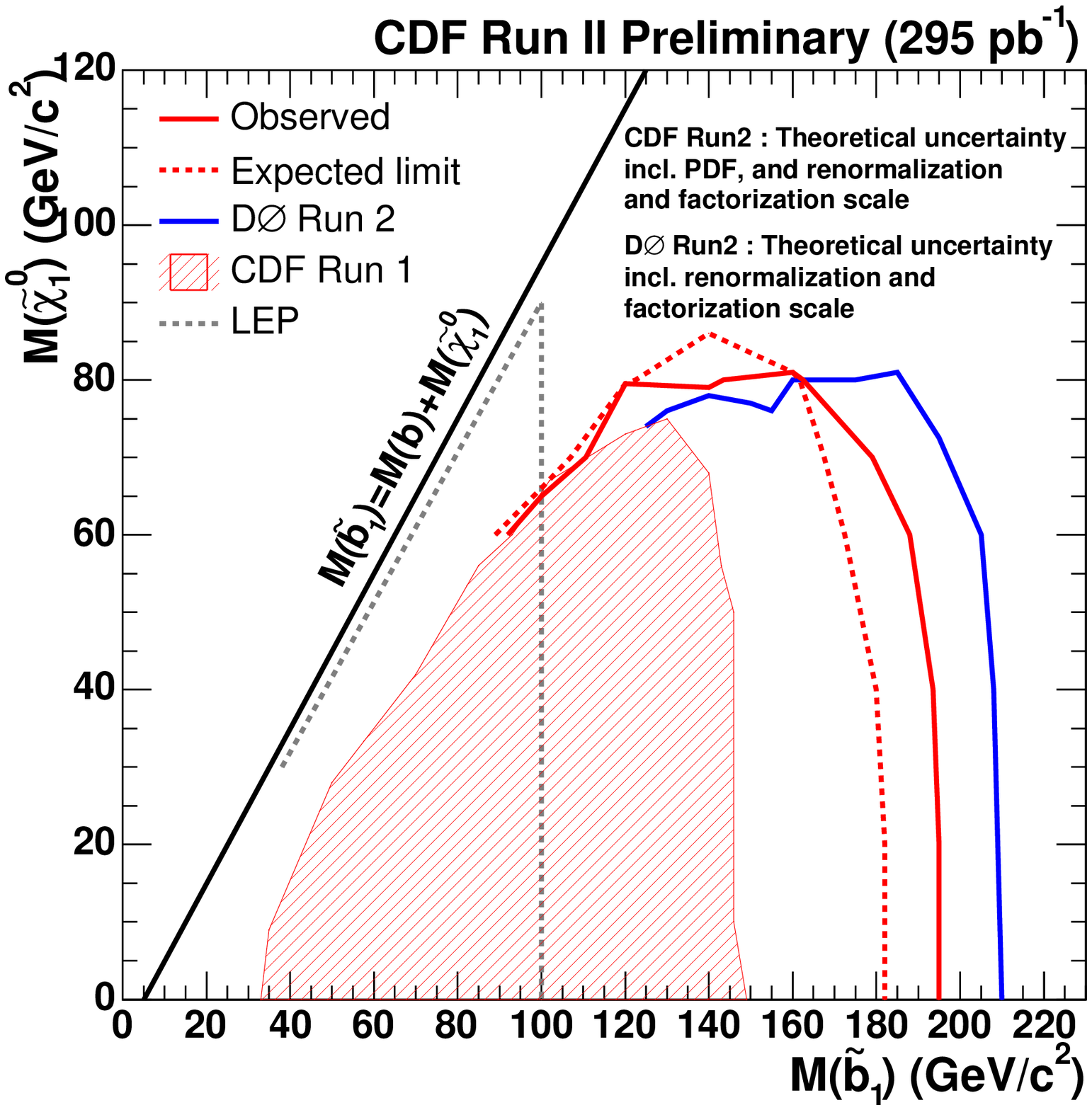}%
 \caption{Limits on neutralino versus stop (left) and sbottom (right) 
   masses from the CDF search in the heavy flavor + \MET\ channel.
   \label{fig:cdfbcjets}}
\end{figure}

CDF has also studied the heavy flavor + \MET\ channel by 
selecting events with at least two jets, one of which must
be heavy flavor tagged~\cite{bib:cdfstopccmet}.  This
channel is interpreted as a search for both
$\tilde{t} \rightarrow c \tilde{\chi}_1^0$ and
$\tilde{b} \rightarrow b \tilde{\chi}_1^0$.
The analysis has three separate
searches based on the hypothetical stop mass ($<$100 GeV,
100-120 GeV, and $>$120 GeV) or sbottom mass ($<$140 GeV,
140-180 GeV, and $>$180 GeV).  Each channel/mass is optimized
and number of data events is compared to the expected
background (see Tab.~\ref{tab:cdfstopccmet}).  Limits in
the neutralino versus stop(sbottom) mass plane are
shown in Fig.~\ref{fig:cdfbcjets}.

\begin{table}[t]
\begin{center}
\caption{Number of data and expected background events for
the CDF heavy flavor + \MET\ analysis.  The left table shows
the stop analysis while the right table is for the sbottom
analysis.}
\begin{tabular}{|c|c|c|} \hline 
 m($\tilde{t}$) & \textbf{Expected Background} & \textbf{Data} \\ \hline
   $<$100 GeV   & $137  \pm 6.2 \pm 14.6$ & 151 \\ \hline
  100-120 GeV   & $94.9 \pm 5.0 \pm 9.9$  & 108 \\ \hline
   $>$120 GeV   & $42.7 \pm 2.6 \pm 4.6$  & 43 \\ \hline 
\end{tabular}
\hspace{2cm}
\begin{tabular}{|c|c|c|} \hline 
 m($\tilde{b}$) & \textbf{Expected Background} & \textbf{Data} \\ \hline
   $<$140 GeV   & $55.0 \pm 4.2 \pm 5.9$ & 60 \\ \hline
  140-180 GeV   & $17.8 \pm 1.7 \pm 1.6$ & 18 \\ \hline
   $>$180 GeV   & $4.7  \pm 2.1 \pm 0.5$ & 3 \\ \hline 
\end{tabular}
\label{tab:cdfstopccmet}
\end{center}
\end{table}

\begin{figure}
 \includegraphics[width=0.55\textwidth]{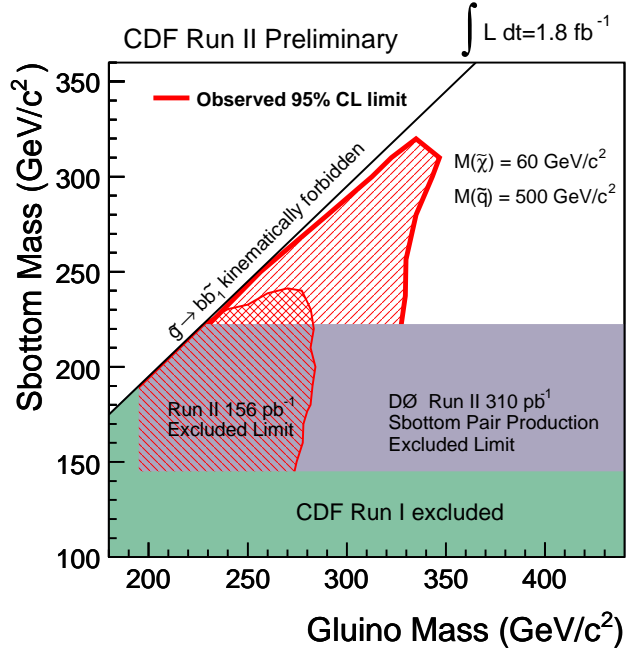}%
 \caption{Limits on the sbottom and gluino masses from the CDF
   search for gluino production.
   \label{fig:cdfgluinosbottom}}
\end{figure}

CDF has also performed a search for gluino production with 
decay mode $\tilde{g} \rightarrow \tilde{b}\bar{b} \rightarrow
b\tilde{\chi}_1^0\bar{b}$~\cite{bib:cdf4bmet}.  For gluino
pair production, this results in a four $b$-jet final state.
The analysis only requires one of the four jets to be tagged
as heavy flavor.  Because the kinematics are heavily 
dependent, the analysis is optimized in two separate regions
of $\Delta m = m(\tilde{g}) - m(\tilde{b})$.  For small
$\Delta m$, 19 events are observed for 22.0 $\pm$ 3.6 
expected while at large $\Delta m$, 25 events are observed
with 22.7 $\pm$ 4.6 expected.  Limits in the sbottom mass
versus gluino mass are shown in Fig.~\ref{fig:cdfgluinosbottom}.

\subsection{Charginos and Neutralinos} 

Associated production of a chargino ($\tilde{\chi}^\pm_1$) and a 
neutralino ($\tilde{\chi}^0_2$) provides a ``golden'' search channel
through events that have three charged leptons and 
missing transverse energy (from neutralinos and neutrinos).  
Very few standard model processes naturally create three
isolated leptons, the exception being dibosons such
as $WZ$.  Other backgrounds tend to be instrumental 
arising from mis-identified isolated leptons.  Due to the
three final state charged leptons and three unobserved
particles, often the lowest $p_T$ lepton will be
difficult to identify cleanly as a lepton.  Therefore,
the search techniques will allow for it to be observed
as an isolated track.  This also allows for some
acceptance of taus that are not explicitly included.

\begin{figure}
 \includegraphics[width=0.45\textwidth]{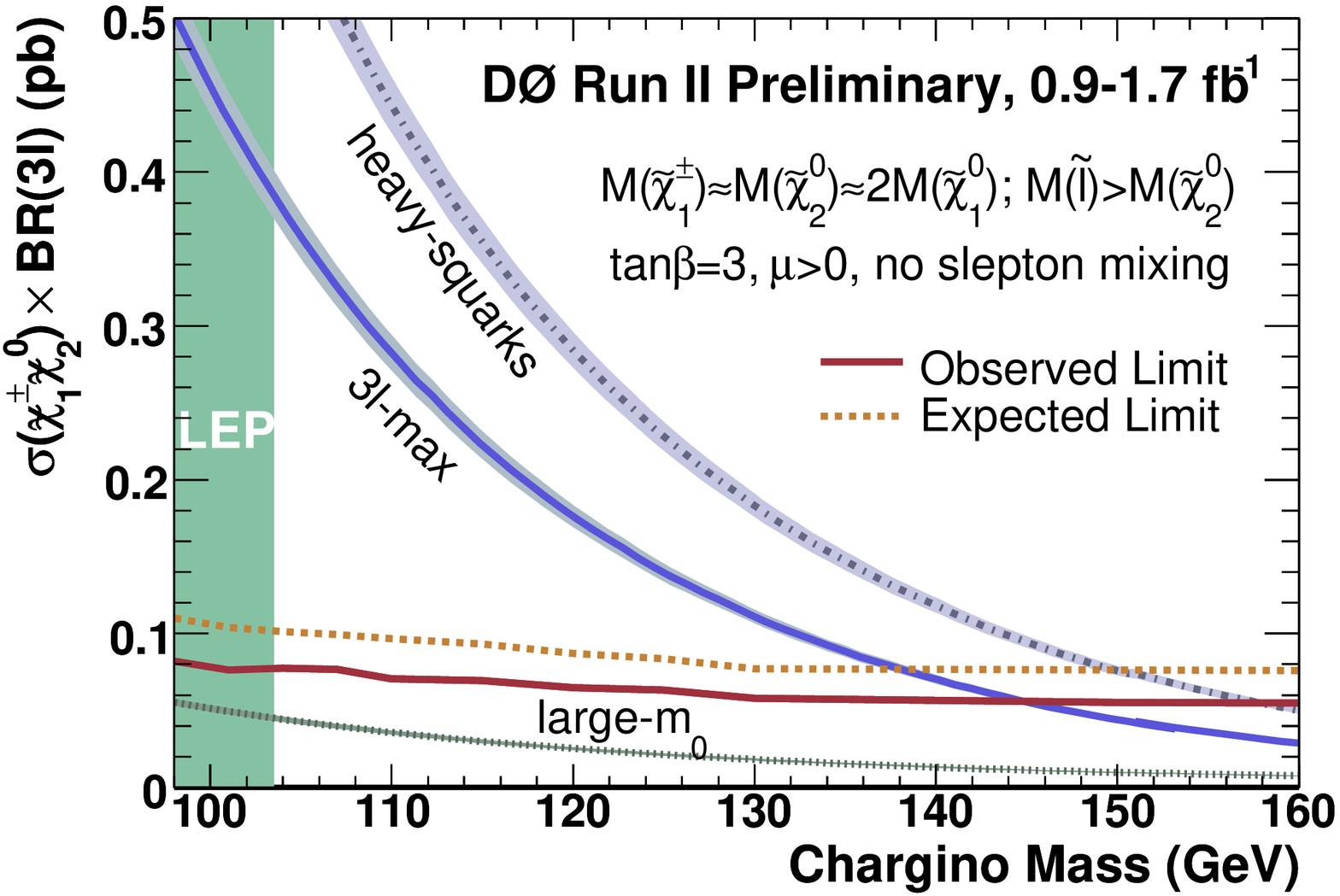}%
 \includegraphics[width=0.45\textwidth]{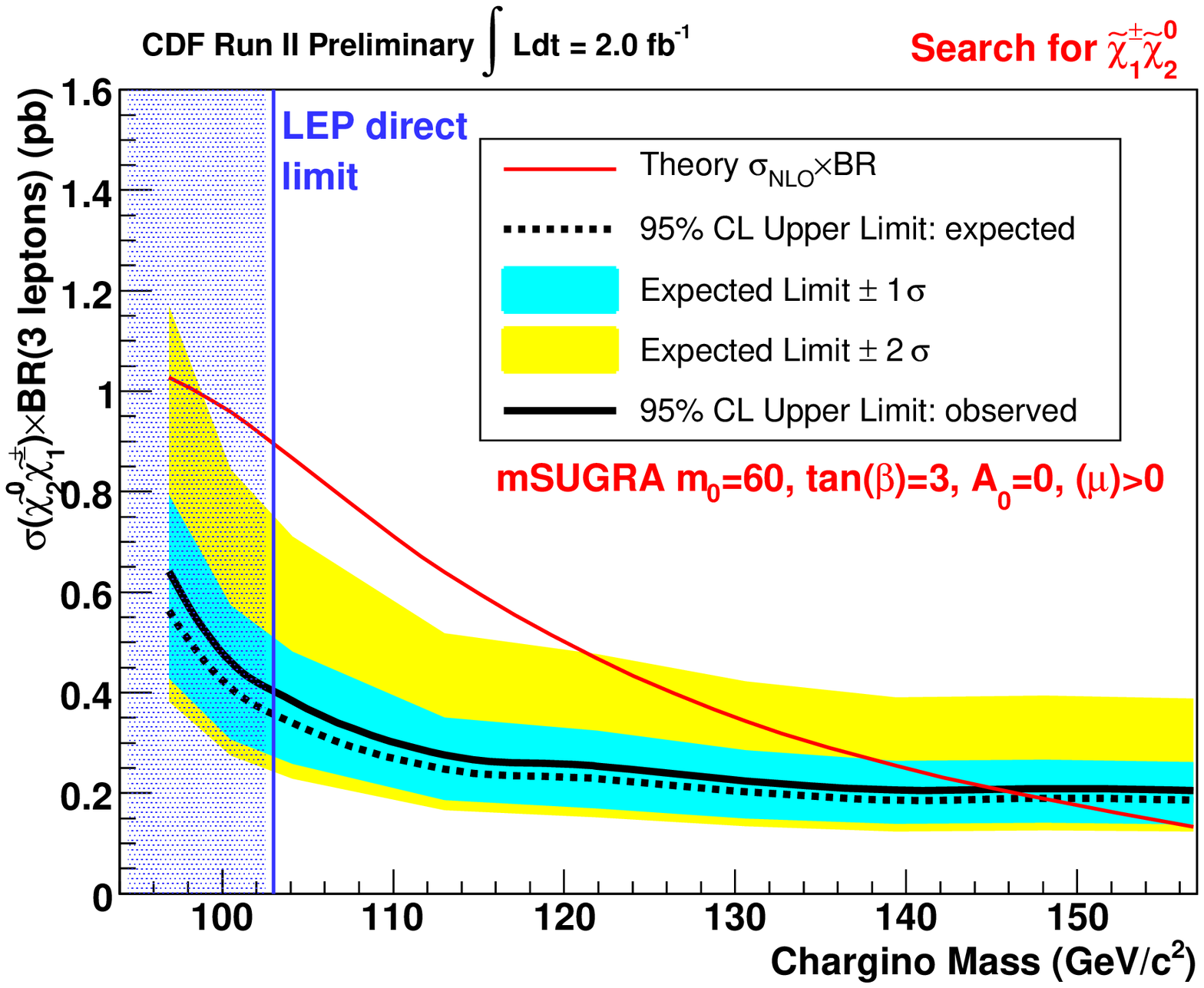}%
 \caption{Limits on the cross section times branching ratio from
   the trilepton analyses of D\O\ (left) and CDF (right). \label{fig:trilep}}
\end{figure}

CDF has performed searches in multiple channels using
2 fb$^{-1}$ of Run II data.  Five exclusive channels 
are ordered by purity according to the quality criteria used
to identify the three constituent leptons. ��They include various
The data and expected backgrounds
are listed in Table~\ref{tab:cdftrilep}.  Excellent
agreement between data and background leads to limits
within the mSUGRA model ($m_0=60$, $tan(\beta)=3$,
$A_0=0$, $\mu>0$) that restrict $m(\chi_1^\pm) > 145$ GeV
(Fig.~\ref{fig:trilep}(left)).  The D\O\ search combines
four analyses with 1-1.7 fb$^{-1}$.
D\O\ categorizes its search by lepton type ($e$, $\mu$ or
track) and also includes a search in same-sign muons
that does not require the third lepton to be observed
(Tab.~\ref{tab:d0trilep}).  D\O\ uses a variation of
mSUGRA with no slepton mixing.  In a version that 
maximizes decays to $e$ and $\mu$ (called 3$\ell$ max),
a limit of $m(\chi_1^\pm) > 145$ GeV is also observed
(Fig.~\ref{fig:trilep}(right)).

\begin{table}[t]
\begin{center}
\caption{Number of data and expected background events for
the CDF trilepton search channels.  The lines labeled ``Total''
are a sum of the lines directly above.}
\begin{tabular}{|c|c|c|} \hline 
 & \textbf{Expected Background} & \textbf{Data} \\ \hline
          3 tight         & $0.49 \pm 0.04 \pm 0.08$ & 1 \\ \hline
     2 tight, 1 loose     & $0.25 \pm 0.03 \pm 0.03$ & 0 \\ \hline
     1 tight, 2 loose     & $0.14 \pm 0.02 \pm 0.02$ & 0 \\ \hline \hline
     Total trilepton      & $0.88 \pm 0.05 \pm 0.13$ & 1 \\ \hline \hline
     2 tight, 1 track     & $3.22 \pm 0.48 \pm 0.53$ & 4 \\ \hline
1 tight, 1 loose, 1 track & $2.28 \pm 0.47 \pm 0.42$ & 2 \\ \hline \hline
  Total dilepton + track  & $5.5  \pm 0.7  \pm 0.9$  & 6 \\ \hline \hline
\end{tabular}
\label{tab:cdftrilep}
\end{center}
\end{table}

\begin{table}[t]
\begin{center}
\caption{Number of data and expected background events for
the D\O\ trilepton search channels.  Here, $\ell$ is an
isolated track.} 
\begin{tabular}{|c|c|c|} \hline 
 & \textbf{Expected Background} & \textbf{Data} \\ \hline
           $ee\ell$           & $1.8 \pm 0.8$         & 0 \\ \hline
         $\mu\mu\ell$         & $0.3 !^{+1.3}_{-0.3}$ & 2 \\ \hline
          $e\mu\ell$          & $0.9 \pm 0.4$         & 0 \\ \hline 
 $\mu^+\mu^+$ or $\mu^-\mu^-$ & $1.1 \pm 0.4$         & 1 \\ \hline 
\end{tabular}
\label{tab:d0trilep}
\end{center}
\end{table}

\begin{figure}
 \includegraphics[width=0.45\textwidth]{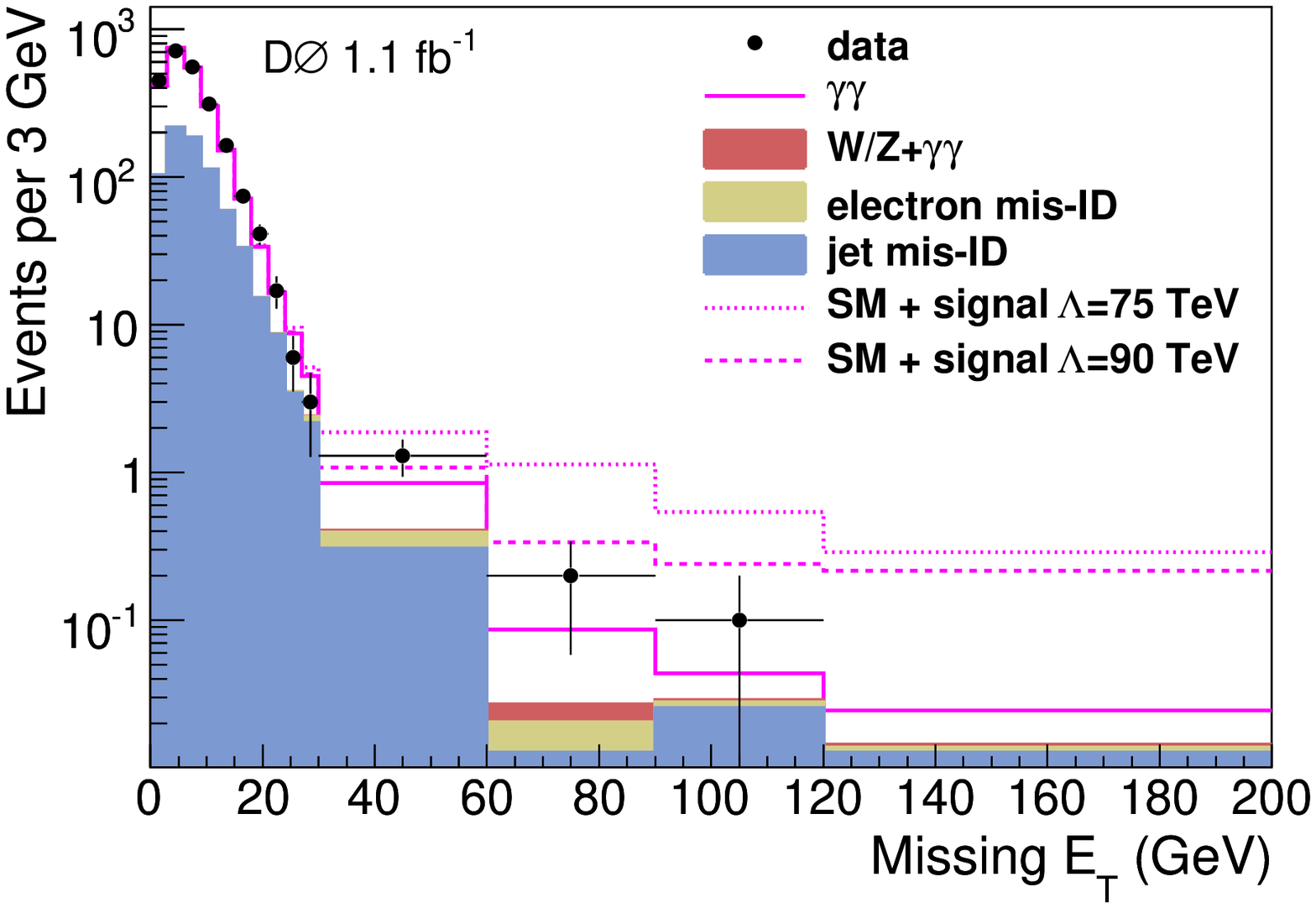}%
 \includegraphics[width=0.45\textwidth]{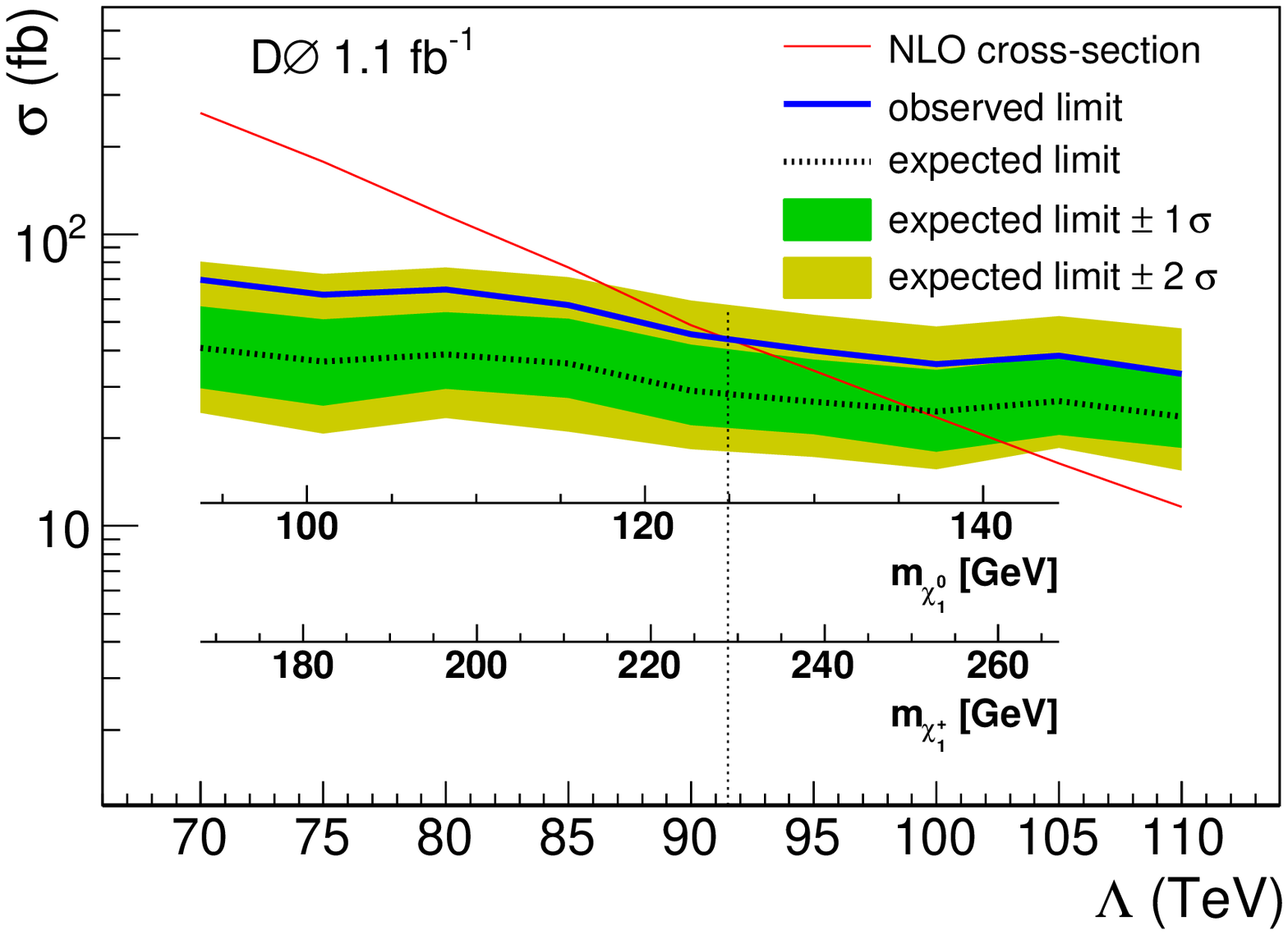}%
 \caption{The \MET\ distribution for D\O\ events with two $p_T > 25$ GeV
   photons (left).  Limits on the production cross section GMSB SUSY
   from the D\O\ di-photon analysis.
   (right). \label{fig:d0gmsb}}
\end{figure}

In the GMSB model, the LSP is the gravitino ($\tilde{G}$) which leads to
a different decay mode $\chi_1^0 \rightarrow \gamma \tilde{G}$ where
the gravitino escapes unobserved.  In events with a pair of
neutralinos, a distinct SUSY signature of two photons plus
large missing transverse energy would be observed.  D\O\
performed a search for such events~\cite{bib:d0gmsb}.  
Figure~\ref{fig:d0gmsb}(left) shows the \MET\ spectrum 
for events with two photons with $E_T > 25$ GeV.  Limits
on the GMSB model from this data result in $\Lambda > 91.5$ TeV,
$m(\tilde{\chi}_1^0) > 125$ GeV, and $m(\tilde{\chi}_1^\pm) > 229$ GeV 
(Fig.~\ref{fig:d0gmsb}(right)).  These
are significant improvements over the previous limits set from
a combination of D\O\ and CDF results with lower
luminosity~\cite{bib:tevgmsb}.  CDF has analyzed the diphoton
channel (see contribution from S.~Yu in these proceedings),
but an interpretation within the GMSB framework is still
in progress.

\section{``UNNATURAL'' SUSY \label{sec:unnatural}}

While the above searches primarily explored traditional variations
of supersymmetry such as mSUGRA and GMSB with \rparity\ conservation,
there are other variations that might exist.  Many of these
will have different experimental signatures that require separate
searches to explore the full parameter space.  Two such categories
have recently been investigated by CDF and D\O: \rparity\
violation and long-lived (but non-stable) particles.  These
are discussed below.

\subsection{R-parity Violating SUSY} 

While conservation of \rparity\ in supersymmetry can allow for a
solution to the dark matter question, there is no {\em a
priori} reason SUSY and dark matter need to be tied
together.  \rparity\ can be trivially violated by adding
terms to the superpotential:
\begin{equation}
 W = W_{MSSM} + \frac{1}{2}\lambda_{ijk} L_i L_j \bar{E}_k
              + \lambda_{ijk}^\prime L_i Q_j \bar{D}_k
              + \lambda_{ijk}^{\prime\prime}\bar{U}_i \bar{D}_j \bar{D}_k
\end{equation}
where the first term is the minimal supersymmetric model (MSSM)
superpotential with \rparity\ conservation, the second and third 
terms allow for lepton number violation,
the fourth allows for baryon number violation, and $i,j,k$ can
have values 1-3 representing the three generations.  It is
common to assume that for each term, only one $i,j,k$
combination dominates.

\begin{figure}
 \includegraphics[width=0.5\textwidth,clip=]{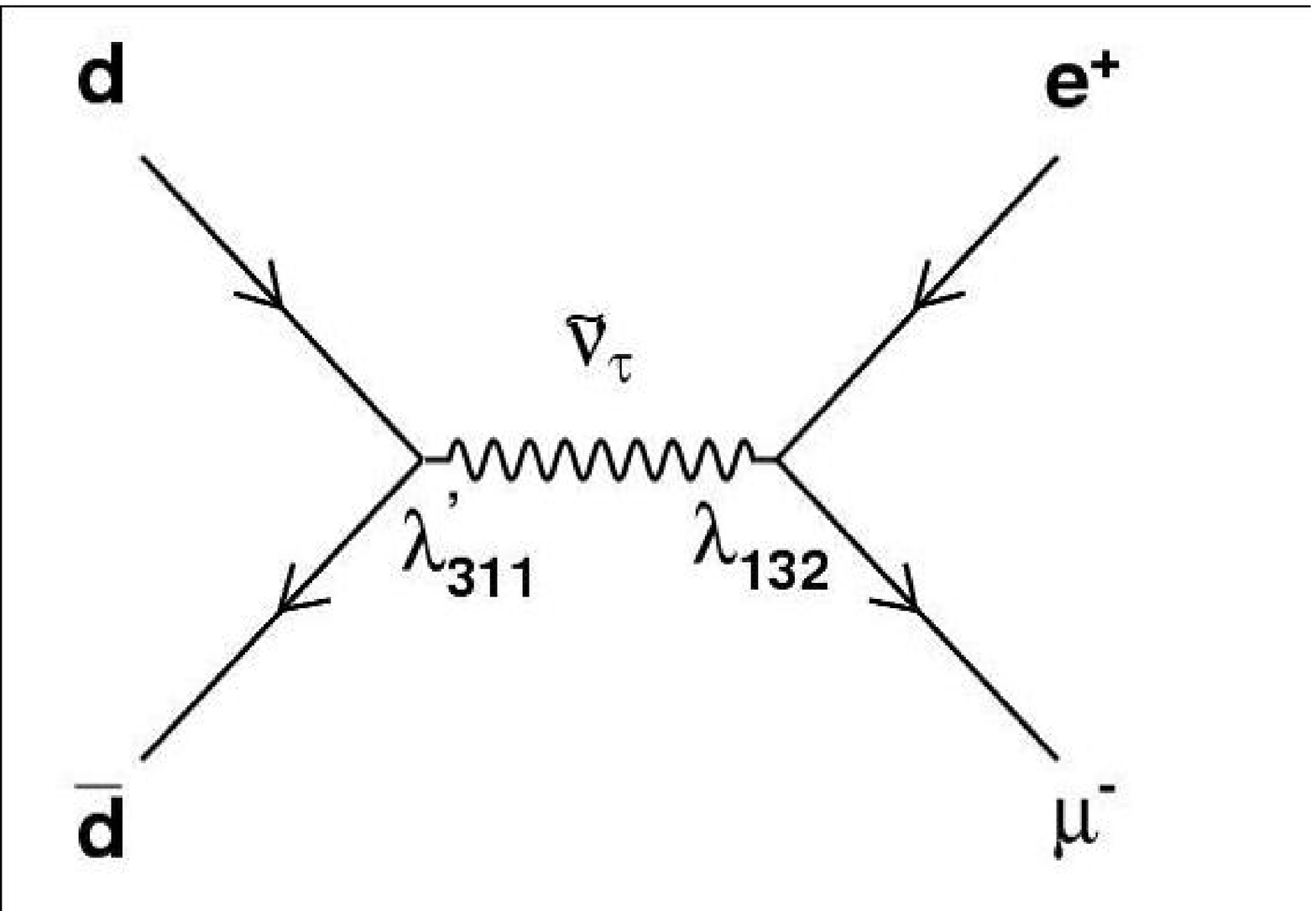}%
 \caption{Feynman diagram for the RPV resonant production of 
   a sneutrino with decay to an electron and a muon.
   \label{fig:snufeyn}}
\end{figure}

\begin{figure}
 \includegraphics[width=0.55\textwidth]{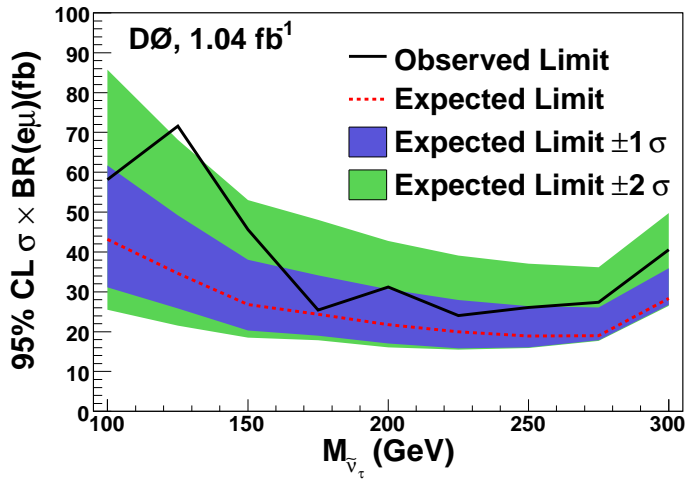}%
 \caption{Limits on sneutrinos
   (right). \label{fig:d0sneutrino}}
\end{figure}

D\O\ has published a search for resonant production of a
sneutrino (Fig.~\ref{fig:snufeyn}) where the production
vertex is governed by $\lambda_{311}^\prime$ and the decay 
vertex by $\lambda_{132}$~\cite{bib:d0sneutrino}.  
In this case, the sneutrino decays to an electron and a muon.  
The $e\mu$ backgrounds are small and arise primarily from 
$Z \rightarrow \tau\tau$, diboson and top quark production.
The search is performed using the invariant mass of the 
two particles where evidence of the sneutrino would
appear as a peak.  Observed data agrees well with 
expected backgrounds and limits on the cross section
times branching ratio are set (Fig.~\ref{fig:d0sneutrino}).

\subsection{Long-lived SUSY Particles} 

Multiple mechanisms exist that can lead to long-lived, but
ultimately unstable, charged or neutral particles in BSM
theories. 
For a review of Run II searches for long-lived particles 
at the Tevatron,
see~\cite{bib:llpreview}.  Recently CDF has performed
two such searches with SUSY interpretations.

\begin{figure}
 \includegraphics[width=0.45\textwidth]{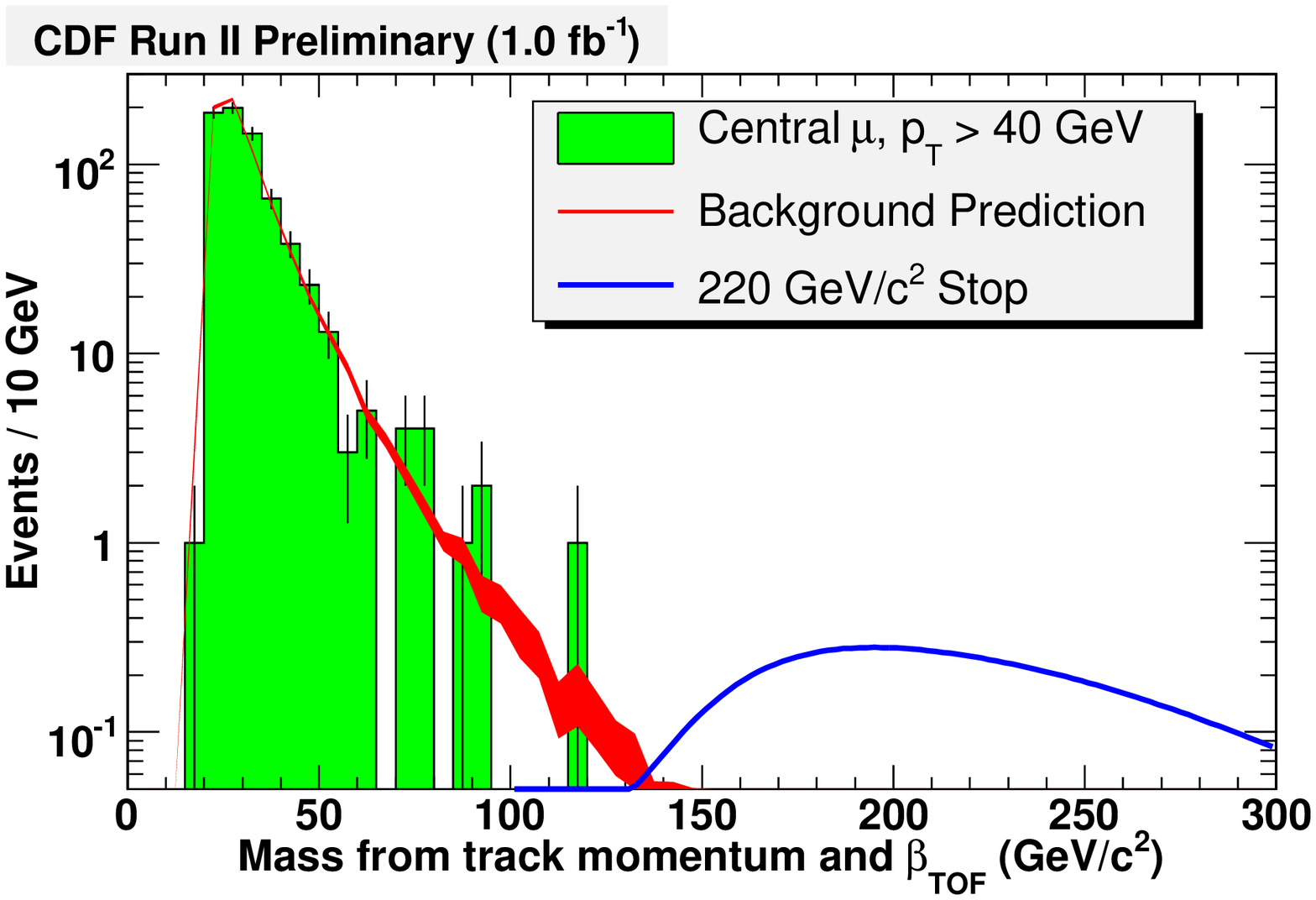}%
 \includegraphics[width=0.45\textwidth]{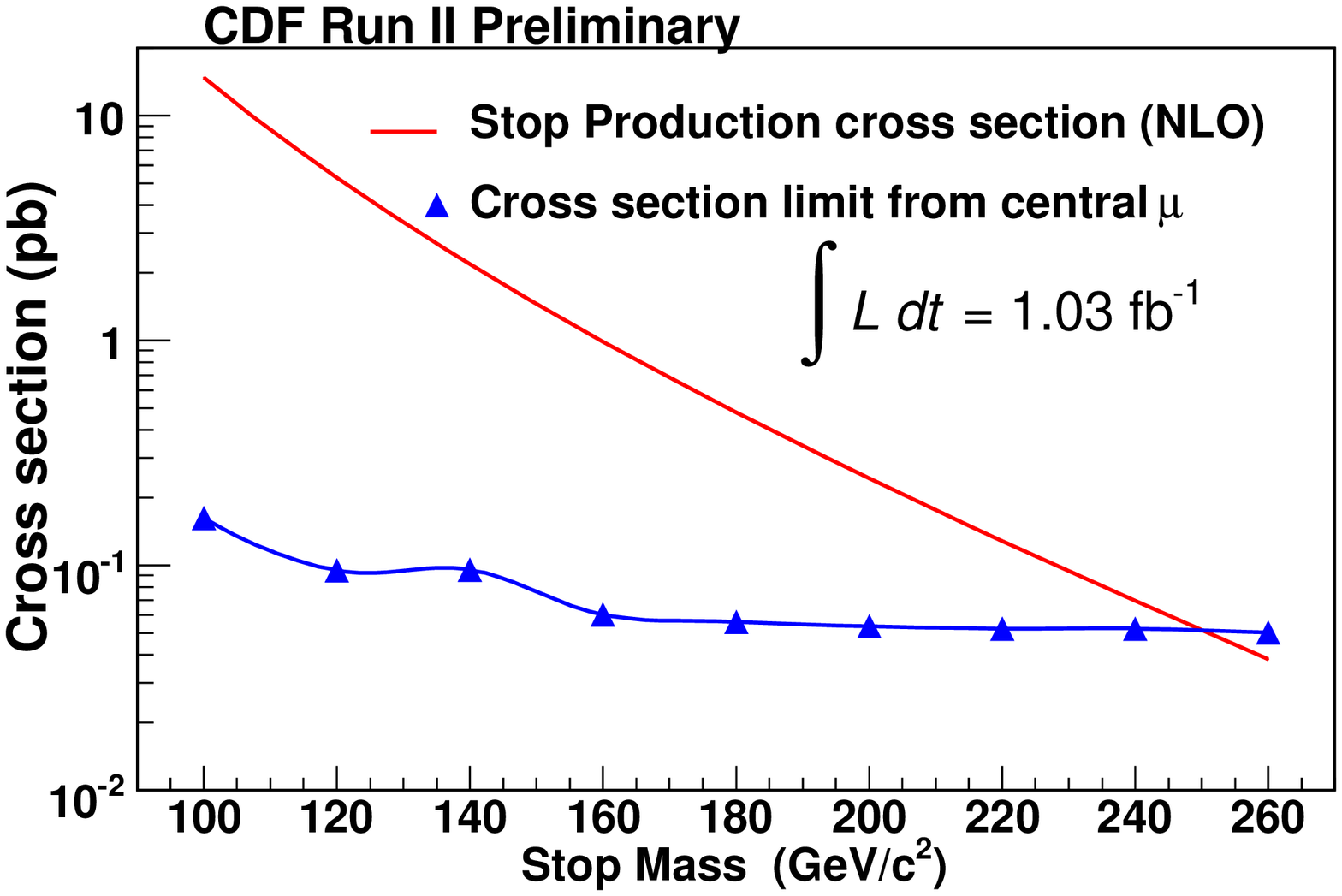}%
 \caption{Spectrum of the mass calculated from the momentum and
   time of flight for CDF muon candidates (left).  CDF limits on the
   production cross section versus stop mass of long-lived stop particles
   (right). \label{fig:cdfllstop}}
\end{figure}

The first search looked for a charged, massive, stable
particle (nicknamed CHAMPs) where stable means 
sufficiently long-lived to escape the detector prior to
decay~\cite{bib:cdfllstop}.  Because the particles 
are massive, they will
tend to move slower than the speed of light and also
will deposit more energy than a minimum ionizing 
particle moving at $c$.
However, they are likely to reach the muon system.
Therefore, CDF searched for particles in the muon
system, measured their time of flight (velocity) and momentum,
and calculated their mass.  Figure~\ref{fig:cdfllstop}(left)
shows the observed mass spectrum.  Since no standard model
CHAMP exists, data is used to estimate the background
to the signal region.  A stable stop that has hadronized
is used as signal and limits on the production cross section
are set (Fig.~\ref{fig:cdfllstop}) which leads to a
stop mass limit of $>$250 GeV.

\begin{figure}
 \includegraphics[width=0.45\textwidth,height=0.4\textwidth]{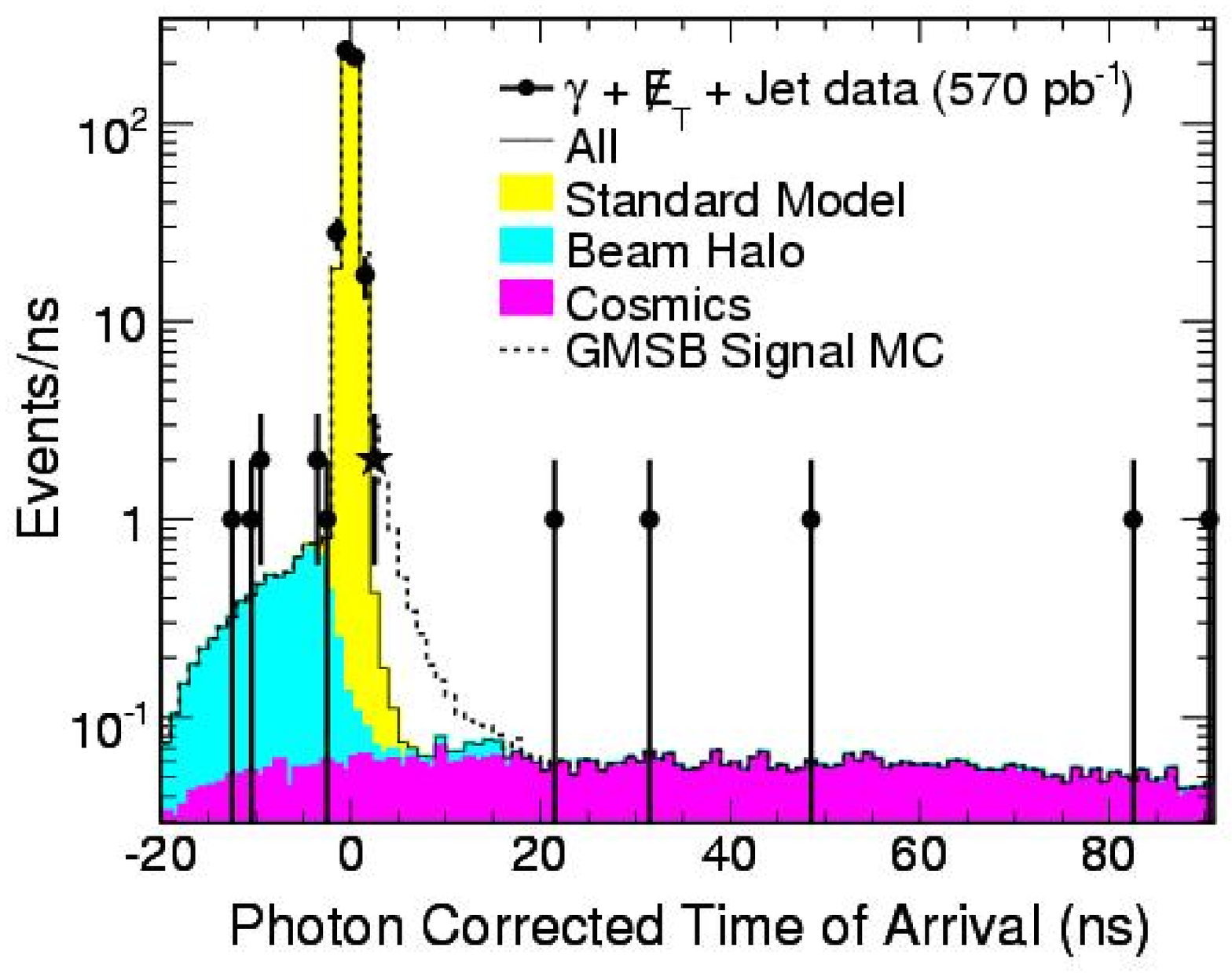}%
 \includegraphics[width=0.45\textwidth,height=0.4\textwidth]{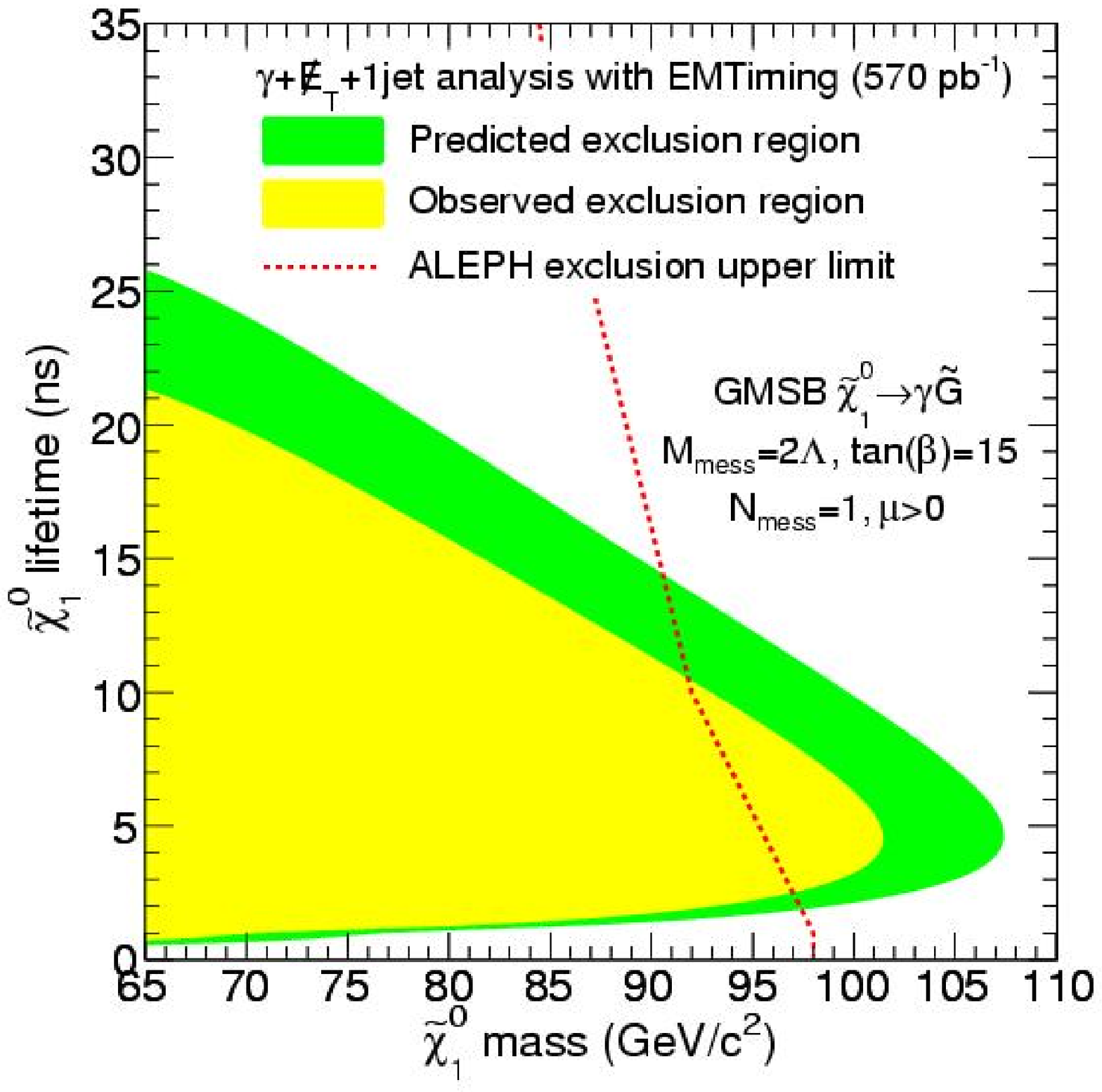}%
 \caption{The difference between the measured and expected arrival
   time of photons in the CDF detector (left).  Standard model
   processes peak around 0.  CDF limits on the neutralino lifetime
   versus mass for long-lived neutralinos (right). \label{fig:cdfllneut}}
\end{figure}

A second search looked for signals from photons that arrive at
the calorimeter later than expected from speed of 
light~\cite{bib:cdfllneut}.  The assumed signal is a long-lived,
slow moving, neutral particle that decays to a photon and an
unobserved particle.  While the search is generally model
independent, a GMSB long-lived 
$\tilde{\chi}_1^0 \rightarrow \gamma \tilde{G}$
model (similar to the GMSB di-photon model above) is used to 
simulate signal.  CDF has instrumented timing for its
electromagnetic calorimeter~\cite{bib:cdfcaltiming}.  
Figure~\ref{fig:cdfllneut}(left) shows the difference between
actual arrival time and expected arrival time for photons.  
Late arriving photons would create an asymmetric tail on the
right side of 0.  In the signal region of 2-10 ns, two
events are observed compared to an expectation of
1.25 $\pm$ 0.66 events.  Limits have been set in the
neutralino lifetime versus mass plane 
(Fig.~\ref{fig:cdfllneut}(right)).

\section{SUMMARY} 

The Tevatron experiments D\O\ and CDF have strong programs to 
search for evidence of supersymmetry.  The analysis techniques
have been well developed and optimized with a good understanding
of the detector and backgrounds.  With larger data sets 
available and being recorded, even more interesting results
are on their way.  In addition, there are expectations that
several results (such a trileptons, squarks and gluinos,
GMSB SUSY) can be combined between the experiments giving
an immediate doubling of the effective luminosity.  While
no evidence of SUSY has been found at the Tevatron, many
of the best available limits have been produced by the
CDF and D\O\ collaborations.

\begin{acknowledgments}
The author would like to thank the CDF and D\O\ collaborations,
particularly members of the exotic and new phenomena groups for
the excellent work that was summarized here.  I'd especially like to 
thank the group conveners, Ben Brau, Chris Hays, and
Patrice Verdier, for their input on the talk and the
proceedings.
\end{acknowledgments}


\end{document}